\documentclass{birkjour}

 \theoremstyle{definition}
 
 \theoremstyle{remark}

 \numberwithin{equation}{section}

\begin{document}

%
%
%
%
%
%
%
%
%

\title[Conformal Inversion and Maxwell Field Invariants]
 {Conformal Inversion\\
 and Maxwell Field Invariants\\
 in Four- and Six-Dimensional\\
 Spacetimes}

\author{Steven Duplij}

\address{%
Dept. of Physics and Technology\\
V. N. Karazin Kharkov National University\\
Svoboda Sq. 4,\\
Kharkov 61077 Ukraine}

\email{Steven.A.Duplij@univer.kharkov.ua, sduplij@gmail.com}

\thanks{Gerald Goldin thanks the organizers of WGMP XXXII for the opportunity to present this talk.}

\author{Gerald A. Goldin}
\address{Departments of Mathematics and Physics\br
Rutgers University\br
Piscataway, NJ 08854 USA}
\email{geraldgoldin@dimacs.rutgers.edu}

\author{Vladimir Shtelen}
\address{Department of Mathematics\br
Rutgers University\br
Piscataway, NJ 08854 USA}
\email{shtelen@math.rutgers.edu}



\date{January 1, 2004}
\dedicatory{Presented in honour of Daniel Sternheimer, on the occasion of his 75th birthday.}

\begin{abstract}
Conformally compactified $(3+1)$-dimensional Minkowski spacetime may be identified with the
 projective light cone in $(4+2)$-dimensional spacetime. In the latter spacetime the special conformal group acts via rotations and boosts, and conformal inversion acts via reflection in a single coordinate. Hexaspherical coordinates facilitate dimensional reduction of Maxwell electromagnetic field strength tensors to $(3+1)$ from $(4+2)$ dimensions. Here we focus on the operation of conformal inversion in different coordinatizations, and write some useful equations. We then write a conformal invariant and a pseudo-invariant in terms of field strengths; the pseudo-invariant in $(4+2)$ dimensions takes a new form. Our results advance the study of general nonlinear conformal-invariant electrodynamics based on nonlinear constitutive equations.
\end{abstract}

\maketitle
\section{Introduction}

Maxwell's equations in $(3+1)$-dimensional spacetime $M^{(4)}$ (Minkowski space) are not only Poincar\'e invariant but conformally invariant. But the physical consequences of this symmetry, if any, remain somewhat unclear.

 As was observed by Dirac \cite{Dirac1935}, the conformal compactification of $M^{(4)}$ (which we denote $M^\#$) can be identified with the projective light cone in a $(4+2)$-dimensional spacetime $Y^{(6)}$, in such a way that the special conformal transformations act by rotations and boosts. One may then write a version of Maxwell's equations in $Y^{(6)}$.

 Introducing so-called hexaspherical coordinates in the latter space, one obtains a spacetime $Q^{(6)}$. Using this coordinatization one seeks to recover classical electrodynamics in $M^{(4)}$ through a process of ``dimensional reduction,'' which involves restriction to the (projective) light cone and the imposition of various conditions on the Maxwell fields. The result is to gain some insight into additional fields that might, as a consequence, survive in $M^{(4)}$. Many details of these results are described by Nikolov and Petrov \cite{NikPet2003}. The conventions we adopt here differ in some ways from their development.

Our first goal in this presentation is to consider how conformal inversion acts explicitly in various coordinate systems. This leads to a number of useful equations. Secondly, we introduce conformal invariant (or pseudoinvariant) functionals of the electromagnetic field strength tensor in $(4+2)$-dimensional spacetime. Our ultimate motivation, in the spirit of our earlier work \cite{GolShtel2001} \cite{GolShtel2004}, \cite{DupGolShtel2008},  is to consider general nonlinear conformal-invariant electrodynamics based on nonlinear constitutive equations. The constitutive equations, in turn, are to be written explicitly in $(4+2)$ dimensions in terms of the conformal-invariant functionals. This allows discussion of both Lagrangian and non-Lagrangian theories. Thus we present here some steps in this overall program.

\section{Maxwell's Equations and Conformal Symmetry}

\subsection{Conformal Transformations of Minkowski space}

We write $ x = (x^\mu) \in M^{(4)}$, with $\mu = 0,1,2,3$. The metric tensor $n_{\mu \nu}$ is $\mathrm{diag}\, [1, -1, -1, -1]$, so that (with the usual summation convention)
$x_\mu x^\mu = n_{\mu \nu} x^\mu x^\nu = (x^0)^2 - (x^1)^2 - (x^2)^2 - (x^3)^2$, and the light cone $L^{(4)}$ is the submanifold $x_\mu x^\mu = 0$. The conformal group then consists respectively of spacetime translations,
\begin{equation}\label{translation}
x^{\prime \, \mu}  =  x^{\,\mu} - b^{\,\mu}\,,
\end{equation}
spatial rotations and Lorentz boosts, e.g.,
\begin{equation}\label{boost}
x^{\prime \, 0} = \gamma (x^0 - \beta x^1)\,, \,\,\, x^{\prime \, 1} = \gamma (x^1 - \beta x^0)\,, \,\,\,
 -1 < \beta = \frac{v}{c} < 1\,, \,\,\, \gamma = (1 - \beta^2)^{-\frac{1}{2}}\,,
 \end{equation}
and dilations,
\begin{equation}\label{dilation}
x^{\prime \, \mu}  =  \lambda x^{\,\mu}\,, \,\,\,  \lambda > 0\,,
\end{equation}
all of which are \textit{causal} in $M^{(4)}$; together with inversion, which breaks causality and acts singularly on the light cone in $M^{(4)}$,
\begin{equation}\label{inversion}
x^{\prime \, \mu}  =  \frac{x^{\,\mu} }{x_\nu x^{\nu}}\,.
\end{equation}
That is, conformal inversion preserves the set of light-like submanifolds (the ``light rays''), but not the causal structure. One may write,
\begin{equation}\label{lightrays}
n_{\mu \nu} dx^{\prime \, \mu}dx^{\prime \, \nu} = \frac{1}{\Omega(x)^2} n_{\mu \nu} dx^{\mu}dx^{ \nu}\,.
\end{equation}
Following inversion by a translation and inverting again gives us the
special conformal transformation,
\begin{equation}\label{specialconformal}
x^{\prime \, \mu}   =  \frac{(x^{\,\mu} - b^{\,\mu} x_\nu x^{\,\nu})}{ (1 - 2b_\nu  x^{\,\nu} + b_\nu b^{\,\nu} x_\mu x^{\,\mu} )}\,\,.
\end{equation}
These can be continuously connected to the identity in the conformal group; thus special conformal symmetry may be studied with (local) Lie algebraic techniques. However, examining the conformal inversion (\ref{inversion}) directly, the main approach taken here, provides valuable insight into the (global) conformal symmetry.

\subsection{Conformal Symmetry of Maxwell's Equations}

Under the transformation (\ref{inversion}), one has the following symmetry transformations of the electromagnetic potential and the spacetime derivatives:
\begin{equation}\label{inversionA}
A_\mu^{\prime}(x^\prime) = x^2 A_\mu(x) - 2x_\mu (x^\alpha A_\alpha(x))\,,
\end{equation}
\begin{equation}\label{inversionpartial}
\partial^\prime_\mu := \frac{\partial}{{\partial x^\prime}^\mu} = x^2 \partial_\mu - 2 x_\mu (x \cdot \partial)\,,
\end{equation}
where we have here used the abbreviations $x^2 = x_\mu x^\mu$ and $(x \cdot \partial) = x^\alpha \partial_\alpha$;
With $F_{\mu \nu} = \partial_\mu A_\nu - \partial_\nu A_\mu$,
\begin{equation}\label{inversionF}
F^{\prime}_{\mu\nu}(x^\prime) = (x^2)^2 F_{\mu\nu}(x) - 2 x^2 x^\alpha (x_\mu F_{\alpha\nu}(x) + x_\nu F_{\mu\alpha}(x))\,,
\end{equation}
and with $\Box = \partial^\mu \partial_\mu$,
\begin{equation}\label{inversionBox}
\Box^{\,\prime} = (x^2)^2 \Box - 4 x^2 (x \cdot \partial)\,.
\end{equation}
Additionally, the $4$-current $j_\mu$ transforms by
\begin{equation}\label{inversionj}
{j^{\,\prime}}_\mu (x^\prime) = (x^2)^3 j_\mu (x)\,-2(x^2)^2 x_\mu (x^\alpha j_\alpha(x))\,.
\end{equation}
These transformations define a symmetry of the (linear) Maxwell equations,
\begin{equation}\label{Maxwell}
\Box A_\nu - \partial_\nu (\partial^\alpha A_\alpha) =  j_\nu\,;
\end{equation}
if $A (x)$ and $j (x)$ satisfy (\ref{Maxwell}), then $A^\prime (x^\prime)$ and $j^{\prime} (x^\prime)$ satisfy the same equation with $\Box^{\,\prime}$ and $\partial^\prime$ in place of $\Box$ and $\partial$ respectively. Combining this symmetry with that of the Poincar\'e transformations and dilations, we have the symmetry with respect to the usual special conformal group.

Note that (\ref{inversionpartial}) and (\ref{inversionBox}) can be obtained by regarding the inversion (\ref{inversion}) as if it were a coordinate transformation, and using the corresponding Jacobian matrix. However (\ref{inversionA}), (\ref{inversionF}), and (\ref{inversionj}) are \textit{symmetry} transformations of the fields, not coordinate transformations.

\subsection{Conformal-Invariant Functionals}

In $M^{(4)}$ we have the Poincar\'e-invariant functionals
\begin{equation}\label{invariantsM}
I_1 \,=\, \frac{1}{2} F_{\mu \nu} (x) F^{\mu \nu} (x)\,,\quad
I_2 \,=\, -\frac{c}{4} {F_{\mu \nu}} (x) {{\tilde{F}}}^{\mu \nu} (x)\,,
\end{equation}
where $\tilde{F}^{\mu \nu} = \frac{1}{2}\,\epsilon^{\mu \nu \rho \sigma} F_{\rho \sigma}$, with $\epsilon$ the usual totally antisymmetric Levi-Civita symbol. Sometimes the functional $I_2$ is called a pseudoinvariant, because it changes sign under spatial reflection (parity). These functionals are useful in writing general nonlinear Poincar\'e-invariant Maxwell systems.

Under conformal inversion, however, $I_1$ and $I_2$ are not individually invariant; rather, they transform by,
\begin{equation}\label{inversionI1}
{I_1^{\prime}} (x^{\prime}) \,=\,  \frac{1}{2} {F_{\mu \nu}^{\prime}} (x^{\prime}) {({F^{\prime})^{\mu \nu}}} (x^{\prime}) \,=\, (x^2)^4 I_1 (x)\,,
\end{equation}
\begin{equation}
{I_2^{\prime}} (x^{\prime}) \,= \,  -\frac{c}{4} {F_{\mu \nu}^{\prime}} (x^{\prime}) {({\tilde{F}}^{\prime})}^{\mu \nu} (x^{\prime}) \,=\, -(x^2)^4 I_2 (x)\,.
\end{equation}
So the ratio $I_2/I_1$ is a pseudoinvariant under conformal inversion. This means, however, that it is invariant under the special conformal transformations.

\section{The Compactification $M^\#$ and the Conformal Group Acting in $(4+2)$-Dimensional Spacetime}

\subsection{Compactified Minkowski Space}

We can also describe Minkowski space using light cone coordinates. Choose a particular (spatial) direction in $\mathbf{R}^3$. Such a direction is specified by a unit vector $\hat{u}$, labeled  (for example) by an appropriate choice of angles in spherical coordinates. A point $\mathbf{x} \in \mathbf{R}^3$ is then labeled by angles and by the coordinate $u$, with $-\infty < u < \infty$, and $\mathbf{x}\cdot \mathbf{x} = u^2$.

With respect to the selected direction, introduce the coordinates
\begin{equation}\label{upm}
u^{\pm} = \frac{1}{\sqrt{2}} (x^0 \pm u)\,.
\end{equation}
Then $\,x_\mu x^\mu \,=\,2u^+ u^-$, so under conformal inversion, with obvious notation,
\begin{equation}\label{inversionu}
u^{\,\prime \,+} \,=\, \frac{1}{2u^-}\,, \quad u^{\,\prime \,-} \,=\, \frac{1}{2u^+}\,\,.
\end{equation}
Now one can compactify $M^{(4)}$ by formally adjoining to it the set $\mathcal{J}$ of the
necessary ``points at infinity.'' These are taken to be the images under inversion of the light cone $L^{(4)}$
(defined by either $u^+ = 0$ or $u^- = 0$), together with the formal limit points of $L^{(4)}$ itself at infinity
(which form an invariant submanifold of $\mathcal{J}$ under conformal inversion). Here $\mathcal{J}$ is the
well-known ``extended light cone at infinity''; see, e.g., \cite{PenRind1984}.

The resulting space
$M^{\#} \,=\,M^{(4)} \cup \mathcal{J}$ has the topology
of $S^3 \times S^1/Z_2$, and conformal inversion acts on $M^\#$ in a well-defined manner.
There are many different ways to coordinatize $M^{\#}$
and to visualize its structure, which we do not review here.

\subsection{The $(4+2)$-Dimensional Space $Y^{(6)}$ and its Projective Light Cone}

One now introduces the $(4+2)$-dimensional spacetime $Y^{(6)}$.
For  $y \in  \mathbf{R}^6$, write $y = (y^m), m = 0, 1, \dots,5$, and define the
flat metric tensor $\eta_{mn} = \mathrm{diag} [1, - 1, - 1, - 1; - 1, 1] $,
so that (with summation convention)
$y_m y^m  \,=\, \eta_{mn} y^m y^n \, =\, (y^0 )^2 - (y^1)^2 - (y^2 )^2 - (y^3 )^2
					 - (y^4)^2 + (y^5)^2$.
The light cone $L^{(6 )}$ is then specified by the condition $y_m y^m  = 0$, or
\begin{equation}\label{lightconeY}
(y^1)^2 + (y^2)^2 + (y^3)^2 + (y^4)^2  \,=\, (y^0)^2 + (y^5)^2\,.
\end{equation}

In  $Y^{(6)}$, define \textit{projective equivalence} in the usual way,
$ (y^m ) \sim (\lambda y^m )$ for $\lambda \in \mathbf{R}, \,\lambda \neq 0$.
The equivalence classes $[y]$ are the rays in $Y^{(6)}$; let  $PY^{(6)}$ denote this space of rays.
The \textit{projective light cone} $PL^{(6)}$ is likewise the space of rays in $L^{(6)}$. To specify
$PL^{(6)}$, one may choose one point from each ray in $L^{(6)}$.
Then, referring back to (\ref{lightconeY}), if we consider
$(y^1)^2 + (y^2)^2 + (y^3)^2 + (y^4)^2  \,=\, (y^0)^2 + (y^5)^2  \,=\, 1$,
we have $S^3 \times S^1$. But evidently we have here \textit{two} points in each ray; so $PL^{(6)}$
can be identified with (and has the topology of)  $S^3 \times S^1/Z_2$.

Furthermore, $PL^{(6)}$ can be identified with $M^{\#}$.
When $y^4 + y^5 \neq 0$, the corresponding element of   $M^{\#}$  belongs to  $M^{(4)}$ (finite Minkowski space), and is given by
\begin{equation}\label{xfromy}
			x^\mu \,=\, \frac{y^\mu}{y^4 + y^5}\,,\quad  \mu \,=\,0,1,2,3\,,
\end{equation}
while the ``light cone at infinity'' corresponds to the submanifold  $y^4 + y^5 = 0$ in $PL^{(6)}$.

\subsection{The Conformal Group Acting in $Y^{(6)}$}

Conformal transformations act in $Y^{(6)}$ via rotations and boosts, so as to leave $PL^{(6)}$  invariant.
We may write this in terms of the $15$ conformal group generators. Setting
$X_{mn} = y_m \partial_n - y_n \partial_m\, (m < n)$,
one has the $6$ rotation and boost generators $M_{mn} = X_{mn}\, (0 \leq m < n \leq 3)$,
the $4$ translation generators
$P_{m} = X_{m5} -  X_{m4}\, (0 \leq m \leq 3)$,
the dilation generator
$D \,=\, - X_{45}$,
and the $4$ special conformal generators,
$K_m \,=\, - X_{m5} - X_{m4}\,  (0 \leq m \leq 3)$.

But of course, from these infinitesimal transformations we can only construct the \textit{special} conformal transformations, which act like (proper) rotations and boosts. Conformal inversion acts in $Y^{(6)}$ by reflection of the $y^5$ axis, which makes it easy to explore in other coordinate systems too:
\begin{equation}\label{inversiony}
{y^\prime}^{\,m} = y^m (m = 0, 1, 2, 3, 4)\,, \quad {y^\prime}^5 = - y^5\,,
\end{equation}
or more succinctly, ${y^\prime}^{\,m} = K^m_n y^n$, where $K^m_n = \mathrm{diag}\,[1,1,1,1,1,-1]$.

\subsection{Maxwell Fields and Conformal Invariants in $Y^{(6)}$}

Now one introduces $6$-component fields $\mathcal{A}_m$ in $Y^{(6)}$, and writes
\begin{equation}\label{Fmn}
\mathcal{F}_{mn} \,=\, \partial_m \mathcal{A}_n - \partial_n \mathcal{A}_m\,,
\end{equation}
so that for any specific choices of $k$, $m$, and $n$,
\begin{equation}
\frac{\partial \mathcal{F}_{mn}}{\partial y^k} + \frac{\partial \mathcal{F}_{nk}}{\partial y^m} + \frac{\partial \mathcal{F}_{km}}{\partial y^n} \,=\, 0\,.
\end{equation}
While this is not really the {\it most\/} general possible ``electromagnetism'' in $(4+2)$-dimensional spacetime, it is the theory most commonly discussed in the linear case. Note that for fields in the space $Y^{(6)}$ we are using the calligraphic font $\mathcal{A}$, $\mathcal{F}$, etc.

To complete Maxwell's equations, we set
\begin{equation}
\frac{\partial \mathcal{G}^{mn}}{\partial y^m} \,=\, \mathcal{J}^{\,n}\,,\label{constitutive}
\end{equation}
where $\mathcal{J}^{\,n}$ is the $6$-current. In the linear theory, $\mathcal{G}$ is proportional to $\mathcal{F}$. For the general nonlinear theory, however, conformal-invariant nonlinear constitutive equations which relate $\mathcal{G}^{mn}$ to $\mathcal{F}_{mn}$ should be written in terms of invariant functionals. Thus the next step is to consider these functionals.

\subsection{Conformal Invariants for Maxwell theory in $Y^{(6)}$}

As we have seen, conformal invariance in $M^\#$ means rotational invariance in $Y^{(6)}$. Thus two rotation-invariant functionals of the field strength tensor $\mathcal{F}_{mn}$ in $Y^{(6)}$ can immediately be written (with $\epsilon$ now the totally antisymmetric Levi-Civita symbol with six indices):
\begin{equation}\label{invariantsY}
\mathcal{I}_1 \,=\, \frac{1}{2} \,\mathcal{F}_{mn}\mathcal{F}^{mn}\,, \quad \,
\mathcal{I}_2 \,=\,\frac{1}{2} \,\epsilon^{mnk\ell rs}\mathcal{F}_{mn}\mathcal{F}_{k\ell}\mathcal{F}_{rs}\,.
\end{equation}
The first rotation invariant functional, perhaps as expected, is analogous to the first invariant in (\ref{invariantsM}) for  the $(3+1)$-dimensional case. But the second rotation invariant functional, unlike the second one in (\ref{invariantsM}), is now \textit{trilinear} in the field strengths (due to the presence of six indices rather than four).

Under conformal inversion, we also have the field transformations,
\begin{equation}\label{AinversionY}
\mathcal{A}^\prime_m (y^\prime) = K^n_m\mathcal{A}_n(y)\,,
\end{equation}
and
\begin{equation}
\mathcal{F}^\prime_{mn} (y^\prime) = - \mathcal{F}_{mn} (y)\,\, \mathrm{if}\,\, m = 5\,\,\mathrm{or} \,\,n = 5\,,\nonumber
\end{equation}
\begin{equation}\label{inversionFmn}
\mathcal{F}^\prime_{mn} (y^\prime) = + \mathcal{F}_{mn} (y)\,\,\mathrm{otherwise}\,.
\end{equation}
So $\mathcal{I}_1$ is invariant under conformal inversion, while $\mathcal{I}_2$ is here seen to be a pseudoinvariant.

\section{Hexaspherical Coordinates and Conformal Inversion in the Space $Q^{(6)}$}

\subsection{Coordinate Transformations}

Hexaspherical coordinates, or $q$-coordinates, are defined conveniently for the eventual process of dimensional reduction. For $q  \in \mathbf{R}^6$, write  $q = (q^a)$, with the index $a = 0,1,2,3,+,- $. Then for  $y \in Y^{(6)}$ with $y^4 + y^5 \neq 0$, define

\begin{equation}\label{defq}
q^\mu  =  \frac{y^\mu}{y^4 + y^5}\,\,  (a = \mu = 0,1,2,3); \,\,\, q^+ =  y^4 + y^5; \,\,\, q^-  = \frac{y_m y^m}{(y^4 + y^5)^2}\,.
\end{equation}
The projective equivalence in $Y^{(6)}$ becomes in $Q^{(6)}$ simply
\begin{equation}
  (q^0, q^1, q^2, q^3, q^+, q^-) \sim (q^0, q^1, q^2, q^3, \lambda q^+, q^-)\,,\quad \lambda \neq 0.
\end{equation}
When we take $q^-$ to zero, we have the light cone in $Q^{(6)}$; when we additional take $q^+ \sim \lambda q^+$, we have the projective light cone and recover Minkowski space.

The inverse coordinate transformation, as well as some later equations, are written more concisely if we introduce the notations
\begin{equation}
(q,q) = (q^0)^2 - \sum_{k=1}^3 (q^k)^2\,, \,\,\, \mathrm{and} \,\,\, Q_{\pm} = (q,q) \pm q^- \,.
\end{equation}
Then
\begin{equation}
y^\mu = q^+ q^\mu \,\, (m = \mu = 0,1,2,3); \,\,\, y^4 =  q^+ \frac{1+Q_{-}}{2}; \,\,\, y^5 = q^{+}\,\frac{1 - Q_{-}}{2}\,.
\end{equation}
The Jacobian matrix for this transformation, defined by
\begin{equation}\label{ydq}
dy^{m}  = \dfrac{\partial y^{m}}{\partial q^{a}}dq^{a}={J}_{a}^{m}\left(
q\right)  dq^{a}\,,
\end{equation}
is given (for rows $m = \mu, 4, 5$; and columns $a = \nu, +, -$) by
\begin{equation}\label{jq}
J_{a}^{m}\left(  q\right)    =\left(
\begin{array}
[c]{ccc}
q^{+}\delta_{\nu}^{\mu} & q^{\mu} & 0\\
q^{+}n_{\nu\sigma}q^{\sigma} & \dfrac{1+Q_{-}}{2} &
-q^{+}/{2}\\
-q^{+}n_{\nu\sigma}q^{\sigma} & \dfrac{1-Q_{-}}{2} &
q^{+}/{2}
\end{array}
\right)\,;
\end{equation}
where $n_{\nu\sigma}=\mathrm{diag}\left[  1,-1,-1,-1\right] $.
The inverse Jacobian matrix expressed in $q$-coordinates, i.e.
$\bar{J}_{m}^{a}\left(  q\right) = J_{m}^{-1,a}\left(  y\left(  q\right)
\right)  $, is then given (for rows $a = \nu, +, -$; and columns $m = \mu, 4, 5$) by
\begin{equation} \label{jq1}
\bar{J}_{m}^{a}\left(  q\right)  =\left(
\begin{array}
[c]{ccc}%
\dfrac{1}{q^{+}}\delta_{\mu}^{\nu} & -q^{\nu}/{q^{+}} & -q^{\nu}/{q^{+}}\\
0 & 1 & 1\\
\dfrac{2n_{\mu\sigma}q^{\sigma}}{q^{+}} & -\dfrac{1+Q_{+}}{q^{+}}
& \dfrac{1-Q_{+}}{q^{+}}
\end{array}
\right) \,.
\end{equation}
In $Q^{(6)}$,  the metric tensor (used to raise or lower indices) is no longer flat. In fact,
\begin{equation}
g_{ab}\left(  q\right)  =J_{a}^{m}\left(  q\right)  \eta_{mn}J_{b}^{n}\left(
q\right)  =\left(
\begin{array}
[c]{ccc}%
\left(  q^{+}\right)  ^{2}n_{\mu\nu} & 0 & 0\\
0 & q^{-} & \dfrac{q^{+}}{2}\\
0 & \dfrac{q^{+}}{2} & 0
\end{array}
\right)  \,,\label{gab}
\end{equation}
while (with raised indices),
\begin{equation}\label{gupperab}
g^{ab}\left(  q\right)  =\left(
\begin{array}
[c]{ccc}%
\dfrac{1}{\left(  q^{+}\right)  ^{2}}\,n_{\mu\nu} & 0 & 0\\
0 & 0 & \dfrac{2}{q^{+}}\\
0 & \dfrac{2}{q^{+}} & - \dfrac{4q^-}{(q^+)^2}
\end{array}
\right)  \,.
\end{equation}
We remark that the coordinate $q^+$ appears explicitly in $\mathrm{det}\,[g^{ab}]\,=\,4/(q^+)^{10} \,=\,(\mathrm{det} \bar{J})^2$, a fact that is important later.

\smallskip
Our next task is to express in $q$-coordinates the invariant functionals $\mathcal{I}_1(y)$ and $\mathcal{I}_2(y)$ given by (\ref{invariantsY}), for which we of course need the field strength tensors in $q$-coordinates. We write the fields $A_a(q)$ and $F_{ab}(q)$ in terms of $\mathcal{A}_m(y)$ and $\mathcal{F}_{mn}(y)$ using the above Jacobian matrices, $A_a(q) = J^m_a (q(y))\mathcal{A}_m(y)$ and $F_{ab}(q) = J^m_a(q(y))\mathcal{F}_{mn}(y)J_b^n(q(y))$.
We have the corresponding inverse transformations,
\begin{equation}\label{inversetransforms}
\mathcal{A}_m (y) = A_a(q)\bar{J}_{m}^{a}( q)\,,
 \quad \mathcal{F}_{mn}(y) = \bar{J}^a_m(q)F_{ab}(q)\bar{J}^b_n(q)\,.
 \end{equation}
 From these equations, it is not hard to demonstrate that $F_{ab}(q) = \partial_a A_b - \partial_b A_a$ (where $\partial_a = \partial / \partial q^a$), using the fact that $\partial_a J^n_b - \partial_b J^n_a = 0$.

In addition, substituting (\ref{inversetransforms}) into (\ref{invariantsY}), one may demonstrate explicitly that in $Q^{(6)}$, the invariants (\ref{invariantsY}) take the form,
 \begin{equation}\nonumber
 I_1(q)= \frac{1}{2}\,F_{ab}(q)F^{ab}(q) = \frac{1}{2}\,g^{ac}g^{bd} F_{ab}(q)F_{cd}(q)\,,
 \end{equation}
 \begin{equation}\label{invariantsQ}
 I_2(q) \,=\, \frac{1}{(q^+)^5}\,\epsilon^{abcdeg}F_{ab}(q)F_{cd}(q)F_{eg}(q)\,\quad
 \end{equation}
 \begin{equation}\nonumber
 \quad \quad =\, \frac{1}{2} (\mathrm{det}\,\bar{J})\,\epsilon^{abcdeg}F_{ab}(q)F_{cd}(q)F_{eg}(q)\,.
 \end{equation}
 Note that $\epsilon$ is the Levi-Civita {\it symbol}. The Levi-Civita {\it tensor} with raised indices is defined generally as $(1/\sqrt{|g|}\,)\epsilon$, where $g = \det [g_{ab}]$. Here this becomes $(\mathrm{det}\,\bar{J})\,\epsilon^{abcdeg}$.

\subsection{Conformal Inversion in $Q^{(6)}$}

The conformal inversion transformation contains most of the essential information
for a subsequent discussion of nonlinear electrodynamics. From (\ref{inversiony}), we obtain the formula for conformal inversion in $Q^{(6)}$,
\begin{equation}\label{inversionq}
{q^{\prime}}^\mu = \frac{q^\mu}{Q_{-}}\,, \quad {q^{\prime}}^{+} = q^{+} Q_{-}\,, \quad {q^\prime}^{-} = \frac{q^{-}}{Q_{-}^{\,2}}\,.
\end{equation}
Recalling that $Q_{-} = (q,q) - q^{-}$, we also have
\begin{equation}\label{inversionQ}
Q_{-}^\prime = \frac{1}{Q_{-}}\,.
\end{equation}

The remaining steps are to express the fields $A^\prime(q^\prime)$ and $F^\prime(q^\prime),$ transformed under conformal inversion, in terms of $A(q)$ and $F(q)$ respectively, and then to explore the dimensional reduction to Minkowski space with attention to the invariants (\ref{invariantsQ}). To do this, we use the conformal inversion of the fields in $Y^{(6)}$ given by (\ref{AinversionY}) and (\ref{inversionFmn}), together with the above Jacobian matrices; for example, $A^\prime_a (q^\prime) = \mathcal{A}^\prime_m (y(q^\prime))J^m_a (q^\prime) =  K^n_m \mathcal{A}_n(y(q))J^m_a (q^\prime)$. The resulting expressions are rather complicated, so we focus here on components especially relevant to the dimensional reduction.

One finds, for example (with $\mu, \nu, \alpha, \sigma = 0, 1,2,3$, and repeated Greek indices summed from $0$ to $3$),
\begin{equation}\nonumber
A^\prime_\nu(q^\prime) = A_\nu(q)\,Q_{-} - 2q^\alpha A_\alpha(q)n_{\nu \sigma}q^\sigma\,\quad\quad
\end{equation}
\begin{equation}\label{AprimeQ}
\quad\quad +\, 2A_+(q)\,q^+ n_{\nu \sigma}q^\sigma\\ - 4A_{-}(q)\,q^{-}n_{\nu \sigma}q^\sigma,
\end{equation}
while
\begin{equation}\nonumber
F^{\,\prime}_{\mu \nu} (q^\prime) = Q_{-}^{\,2}\,F_{\mu \nu} - 2Q_{-}\,q^\alpha\,(q_\mu F_{\alpha \nu} + q_\nu\,F_{\mu \alpha})\,\quad \quad\quad
\end{equation}
\begin{equation}\label{FprimeQ}
 \quad \quad \quad \quad\quad \,+\,\, \mathrm{terms\,in\,other\,components\,of\,} F.
\end{equation}

\section{Remarks on the Conformal Invariants and Dimensional Reduction}

Note that if $q^- \to 0$, then $Q_{-} \to (q,q)$, and (\ref{inversionq}) becomes
\begin{equation}
{q^{\,\prime}}^\mu = \frac{q^\mu}{(q,q)}\,, \quad {q^{\prime}}^{+} = q^{+} (q,q)\,, \quad {q^\prime}^{-} = 0\,.
\end{equation}
Thus when we move to the light cone in $Q^{(6)}$, identifying the first four components $q^\mu\,(\mu = 0,1,2,3)$ with the point $x = (x^\mu) \in M^{(4)}$ and identifying $(q,q)$ with $x_\mu x^\mu$, we recover the formula (\ref{inversion}) for conformal inversion in $M^{(4)}$.

The condition $q^- = 0$ is preserved by conformal inversion, as is the equivalence relation $(q^\mu, q^+,0) \sim (q^\mu, \lambda q^+, 0), \lambda \neq 0$. However, note that the prescription $q^+ = 1$ for selecting a particular element of each equivalence class is \textit{not} invariant under conformal inversion.

Now it is instructive to compare (\ref{FprimeQ}) with the corresponding expression (\ref{inversionF})  in $M^{(4)}$ for $F^\prime_{\mu \nu}(x^\prime)$; the two are formally the same (up to the terms included) when $Q_{-}$ is taken to $(q,q) = q_\rho q^\rho$. However, $I_1(q) = (1/2)F^\prime_{ab} (q) {F^\prime}^{\,ab} (q)$ defines an invariant under conformal inversion. In contrast, $I_1(x) = (1/2) F^\prime_{\mu \nu} (x) {F^\prime}^{\,\mu \nu} (x)$ transforms according to (\ref{inversionI1}) and is not invariant.

The reason for this difference is now clear. The metric tensor $g$ in $Q^{(6)}$, given by (\ref{gupperab}), is applied twice to raise the indices $a$ and $b$ in the expression for $I_1(q)$. This introduces an additional factor of $1/(q^{+})^4$ as compared with the corresponding expression for $I_1(x)$ in $M^{(4)}$. Under conformal inversion, ${q^\prime}^{\,+} = q^+Q_{-}$, which reduces to $q^+\,(q,q)$ when $q^- \to 0$. When we then identify $q^\mu$ with the coordinates of Minkowski spacetime, the resulting fourth power of $q_\mu q^\mu$ in the denominator restores the invariance under conformal inversion.

 Evidently the dimensional reduction procedure for conformal invariant nonlinear Maxwell theories in $(4+2)$-dimensional spacetime, with compactified Minkowski space identified with the projective light cone, must take account of the fact that setting $q^+ = 1$ (as a device for handling the projective equivalence) is inconsistent with the desired conformal symmetry. This is important if we are to write nonlinear constitutive equations in terms of the $(4+2)$-dimensional invariants.

\subsection*{Acknowledgments}
SD acknowledges the support of a Fulbright Fellowship for the academic year 2011-12 at Rutgers University, where some of this research was conducted. GG also thanks Rutgers University for travel support.

\end{document}